\documentstyle[11pt]{article}
\def\tts{\small\tt }

\begin{document}
\vspace*{-17mm}
\hfill Data Science Journal, Volume 8 (2009), in press  (http://dsj.codataweb.org) \\[1mm]

{\large\bf THE CATS SERVICE: AN ASTROPHYSICAL RESEARCH TOOL}

\textrm{\bfseries\slshape O.V. Verkhodanov\,$^1$, S.A. Trushkin\,$^1$, H. Andernach\,$^2$,
and V.N. Chernenkov\,$^1$}

{\it $^1$\,Special Astrophysical Observatory, Nizhnij Arkhyz, Karachaj-Cherkesia,
  Russia \\
\hspace*{2mm} E-mail:} \underline{vo@sao.ru}, \underline{satr@sao.ru}, \underline{vch@sao.ru} \\
{\it $^2$\,Argelander-Institut f\"ur Astronomie, Bonn, Germany \\
\hspace*{2mm} (on leave of absence from Departamento de Astronomia,
Universidad de Guanajuato, Mexico) \\
\hspace*{2mm}  E-mail:} \underline{heinz@astro.ugto.mx} \\[2ex]

\centerline{\textrm{\bfseries\slshape ABSTRACT}}

{\it
We describe the current status of CATS (astrophysical CATalogs
Support system), a publicly accessible tool maintained at Special
Astrophysical Observatory of the Russian Academy of Sciences (SAO RAS)
(http://cats.sao.ru) allowing one to search hundreds of catalogs
of astronomical objects discovered all along the electromagnetic
spectrum. Our emphasis is mainly on catalogs of radio continuum sources
observed from 10 MHz to 245 GHz, and secondly on catalogs of objects
such as radio and active stars, X-ray binaries, planetary nebulae, HII
regions, supernova remnants, pulsars, nearby and radio galaxies, AGN and
quasars. CATS also includes the catalogs from the largest extragalactic
surveys with non-radio waves. In 2008 CATS comprised a total of about
10\,$^9$ records from over 400 catalogs in the radio, IR, optical and X-ray
windows, including most source catalogs deriving from observations with
the Russian radio telescope RATAN-600. CATS offers several search tools
through different ways of access, e.g. via web interface and e-mail. Since
its creation in 1997 CATS has managed about 10\,$^5$ requests. Currently CATS
is used by external users about 1500 times per day and since its opening
to the public in 1997 has received about 4000 requests for its selection
and matching tasks.
}


\bigskip
{\large\bf 1 INTRODUCTION}

Vast amounts of astrophysical information are now being published,
based on observations of small and large sky regions. Typically, this
information includes coordinates of the observed objects and their
physical characteristics in the form of source catalogs. In fact,
almost every new large-scale observational experiment produces a new
catalog of objects. Modern astrophysics operates with source parameters
obtained in different wavelength ranges with the goal of obtaining the
most detailed understanding of physical properties and the processes
of radiation of these objects. The ability of using different catalogs
makes this problem considerably simpler.

Over the past decades several different attempts have been made to
combine large numbers of astronomical catalogs and make them accessible
in a unified way, which can be classified grossly into two categories:
databases of objects and catalog browsers. Examples of the former
are NED (Helou et al., 1990; Mazzarella et al., 2002), Simbad (Egret,
1983; Wenger et al., 2000), and LEDA (Paturel et al., 1997). Examples
of the latter are Vizier (Ochsenbein et al., 2000) and Datascope
(http://heasarc.gsfc.nasa.gov/vo/).

Motivated by RATAN-600 observation requirements, problems of radio source
study, and the underrepresentation of radio source catalogs in the
then existing object databases, the present authors decided, in 1995,
to create CATS, the astrophysical CATalogues Support system. Following
the radioastronomical needs, the architecture of the operating system
and the considerations expressed in the reviews by Andernach (1990,
1994, 1999), we chose to design CATS as a catalog browser rather
than an object database, given that it would allow us to achieve a
much better completeness in terms of the number of records, implying
e.g. a more complete coverage of radio source measurements over the
entire radio frequency window. We thus deliberately delayed the known
problem of cross-identification of all the catalogs, as provided in
NED, SIMBAD, or LEDA, to a later stage. In that sense CATS is ideally
suited for the experienced researcher who is looking for the largest
amount of data available, but willing and able to work out the correct
cross-identifications by himself. This effort will be compensated for
by more complete data than those obtained from other existing object
databases.

The first steps of the creation of CATS were described by Verkhodanov and
Trushkin (1994, 1995a,b), Verkhodanov et al.\ (1997, 2000), and Trushkin
et al.\ (2000). CATS allows one to operate with catalogs stored in ASCII
and FITS binary tables, to fit radio continuum spectra, and to calculate
spectral indices obtained from different radio catalogs. This service
is located on the server 'cats.sao.ru' at SAO, Russia, and currently
runs on a Dual Opteron 244 under OS Linux Fedora Core 6.

\bigskip
{\large\bf 2~~~~~~~IMPLEMENTATION OF CATS}

There are many modern databases operating with standard software that
allow a user to obtain information on celestial objects, their images
at different wavelengths, and the corresponding bibliographic data.
The above-mentioned services like NED, SIMBAD, and LEDA are among
these. Other services, which may be called ``virtual telescopes",
offer extractions from large, dedicated imaging surveys, such
as the Sloan Digital Sky Survey (SDSS) at www.sdss.org, 2MASS at
www.ipac.caltech.edu/2mass/, and SkyView at skyview.gsfc.nasa.gov
make use of modern Web-technologies allowing one to select sky regions
either via coordinates or object name resolvers. We developed our own
simple approach, following both the RATAN-600 observational tasks for
the preparation of radio source spectra, as well as users' desire to
be able to select observing targets from available source catalogs.
Based on simple system commands and basic functions in C, our approach
permitted us to achieve the fast speed of source selection and easy
portability among different Unix-like systems. Moreover, this approach
was better suited to our very limited resources as it allowed easier
integration of new catalogs. We believe that our best contribution is
a larger number of searchable catalogs rather than a more sophisticated
search interface with less accessible data. We also wish to emphasize that
we are not trying to simply duplicate other catalog browsers by using
their sets of catalogs. A major fraction of these has been obtained or
prepared by members of our team (see e.g. Andernach 2009), and over 100
radio source lists available from CATS are not available from VizieR.

The present data collection consists of catalogs, their descriptions,
and corresponding programs operating under OS Linux. The program codes
are created in the C language and translated with the GNU project
C compiler. The codes are freely shared, provided they are used for
non-commercial goals.

New catalogs may be added to the system in conformity with the
following rules: \\
1) Every new catalog of objects should be placed in the UNIX directory
with the same name as the catalog of objects; \\
2) The description of the catalog should also be placed in that directory; \\
3) The programs (or operating scripts) for local operations with the
catalog of objects are also placed in the same directory; and \\
4) Brief characteristics, program names and description file of the
astrophysical catalog are stored in a file named 'cats\_descr'.

The following information is stored in the description file 'cats\_descr': \\[1mm]
$\bullet$ the name of the catalog, which coincides with the name of the UNIX
directory,  \\
$\bullet$ the type of the catalog (radio, optical, combined, etc.), \\
$\bullet$ frequency/wavelength range,  \\
$\bullet$ minimum fluxes or magnitudes,  \\
$\bullet$ equatorial or/and galactic coordinate ranges of the repective catalog, \\
$\bullet$ names of the local programs for the 'select' and 'match' functions
(see Section~3.2), \\
$\bullet$ name of the document file,  \\
$\bullet$ number of records in the catalog,  \\
$\bullet$ the size of the beam pattern or angular resolution,  \\
$\bullet$ a recalibration factor (if available) to put the flux densities on
a commonly agreed flux scale, and  \\
$\bullet$ a reference.

Parameters from the description file are used by the programs that process
the input data, e.g., user-specified limits in coordinates and/or flux
density inform the operating programs about the relevant sky zones to
be searched and thus economize on processing time if a given catalog is
out of the required range.

The lower level of the CATS control system includes several basic
utilities: \\
(1) 'c\_sel' selects objects with parameters within the user-specified
ranges; \\
(2) 'c\_match' looks for objects falling within a certain distance from
a user-specified coordinate (a ``cone search"); \\
(3) 'cats\_divide' operates with catalogs sorted by right ascension and
produces an index of record numbers corresponding to a certain right
ascension; \\
(4) 'epoch' converts the user-specified coordinates  if they are provided
for non-standard equinox, i.e. neither 1950.0 nor 2000.0 (proper motions
are assumed to be zero); \\
(5) catalog coordinates stored according  to the equinox as published,
and user-specified coordinates are converted to that equinox; \\
(6) 'cats\_base' controls the file of descriptions 'cats\_descr' selecting
characteristics of catalogs and produces input parameters for the programs
'c\_sel' and 'c\_match.'

For catalogs in FITS binary table format (e.g., NVSS or SDSS catalogs),
special programs for the 'select' and 'match' functions were prepared.

To organize access of a local user of CATS from any directory of the OS,
the operation programs that process the main CATS tasks are placed in
a commonly accessible directory of the OS Linux. The control procedures
'cats\_sel' and 'cats\_match,' organized as shell-scripts, cover all the
low-level programs and provide the interface between CATS and a local
user, or requests made via HTTP or e-mail. The described method of the
catalog storage facilitates the database development, its expansion with
new data, and the possibility to tune the supporting programs.

CATS has its own indexing procedure 'cats\_divide' for object coordinates
in a catalog. The program decides where to start searching records of the
input catalog from an index prepared from a right-ascension sorted list,
allowing one to make efficient use of hard-disk seeking functions. To
avoid problems of searching within a few degrees of the poles where
recalculation from one equinox to another one is done, we process the
total list of objects in these zones to find a required source. The
indexing of CATS lists by declination is now under consideration.

\bigskip
{\large\bf 3~~~~~~~~THE MAIN FUNCTIONS}

The following functions are currently implemented in CATS: \\[1mm]
(1) Selection of objects from one or several catalogs by the following
criteria: equatorial or galactic coordinates, flux densities, spectral
indices, frequencies, names, and (for some catalogs)  the type of
objects. Parameters common to all catalogs (coordinates and flux density)
are used in the selection. \\
(2) Search for counterparts of objects (selected from one or several
catalogs) by coordinate matching within a box, a circle, or an ellipse. \\
(3) Cross-identification of different catalogs. This procedure is
currently available only for local users. It uses the output from the
selected zones of one catalog as input for the matching procedure on
another catalog. \\
(4) Preparation of a file with a short description and characteristics
of each catalog, printing of the total list of catalogs for the required
sky areas (for local user). \\
(5) Plotting radio spectra of selected sources. This can be done from
multi-frequency catalogs, or catalogs prepared by the CATS team from
cross-identification of radio catalogs at different frequencies, or even
by individual user input of frequencies and flux densities for a source
(at \verb*Ccats.sao.ru/~satr/SP/spectrum.htmlC). This function is realized in two
procedures. The first is a Java script for data in the homogenized CATS
format for single objects, accessible from the web page. The second one is
suitable for object lists in the local RATAN-600 data processing system.

The result of the object selection is a datafile sorted according to
different object characteristics, such as right ascension (default
mode), declination or frequency. This file can either be displayed on
the standard output devices or sent to the user in the following formats: \\
(1) The 'native' (original) format of the catalog (i.e. columns as
published). \\
(2) Standard (``homogenized") output format. This format is organized to
be common for all the catalogs and used later for unification of data,
preparation of spectra and statistical analysis. The standard FITS Table
format describing data and fields of the table may be added as a header
of the resulting file.

\bigskip
{\large\bf 3.1~~~~~~~Access to CATS}

Different modes of access are provided according to user requirements
and CATS goals, following modern trends of software development. Thus,
three main modes of on-line access to the CATS service have been prepared: \\
(1) Dialog mode (non-graphics) has been maintained until the
present. Several scripts written in the UNIX 'shell' language have
been created (Verkhodanov and Trushkin 1994). They permit one to operate
with CATS' supporting programs via TCP/IP and NFS protocols in the local
computer network. \\
(2) Hypertext access (http://cats.sao.ru) is provided to allow a user
from the Internet system to operate with CATS via a hypertext transfer
protocol. It allows one to execute all described operations from the
Web-page. \\
(3) FTP access (ftp://cats.sao.ru) allows a user to obtain both the
description of CATS catalogs and the catalogs themselves. Here is an
example of access by anonymous FTP to the catalog WISH: \\[-11mm]
{\small
\begin{verbatim}
            anonymous@ftp://cats.sao.ru
            ftp>  bin
            ftp>  cd pub/cats/
            ftp>  cd  WISH
            ftp>  get wish11.cat
            ftp>  bye
\end{verbatim}
}

\vspace*{-5mm}
All the catalog names are stored in a file of descriptions 'cats\_descr'
(ftp://cats.sao.ru/cats\_descr). \\
(4) E-mail access allows the user to send batch requests to CATS.

One can send an e-mail with requests of matching with lists or selection
by several parameters. The e-mail will be read automatically and sent for
execution to the CATS scripts. The result will be sent back automatically
to the user via e-mail.

\bigskip
{\large\bf 3.2~~~~~~'Select' and 'match' procedures}

Two main procedures of data selection have been provided at different
levels of CATS control: the 'select' and 'match' tasks. They follow the
first three main functions described above and are realized in different
access modes. As was described earlier, the two low-level procedures
'c\_sel' and c\_match' provide information according to the corresponding
user requests and pass it to the upper-level control scripts 'cats\_sel'
and 'cats\_match'. Using these two procedures, the cross-identification
of catalogs is possible by using output from the `select' function of
the first complete catalog as the input of the `match' function of the
second catalog or catalogs.

\bigskip
{\large\bf 3.3~~~~~~E-mail access}

In order to economize user's real time and avoid delays in on-line data
exchange, we provide the possibility to submit a batch task in the form
of an e-mail letter. The e-mail requests may have several formats (as
explained in a file that is returned on an empty e-mail to cats@sao.ru,
with no subject). Two examples of the body of an e-mail (no subject
required) are shown below. \\[-3ex]

(1) selection task: ~~~~~~~~~~~~~~~~~~~~~~~~~~~~~~~~~~~ (2) matching task: \\[-7mm]
{\small
\begin{verbatim}
 mail -s "" cats@sao.ru                   mail -s "" cats@sao.ru
     cats select                              cats match
     ra min=10:30 max=10:40:00.               catalogs NVSS,FIRST
     dec > 10' < 12' 30"                      window box x=30" y=10'
     flux > 0.5mJy                            sources:
     catalogs r epoch=1950                    s1 02:02:00    +31:23:16   1950
     cats end                                 s2 02:23:10     34:03:00   1950
                                              s3 21:26:33.9  -18:34:33.0 1950
                                              cats end
\end{verbatim}
}

These examples demonstrate the use of some keywords for batch
requests. The beginning and end of the requests are defined by the
statements: 'cats start' and 'cats end'. The first example shows how to
search within certain limits of coordinates by using the 'min', 'max' or
'$<$', '$>$' operators for right ascension 'ra' and declination 'dec'. The
keywords expression 'catalogs r' sets the type of catalogs to be used
for selection, where 'r' means all the radio catalogs. Instead of
'r', one may choose 'o' for the optical catalogs, 'x' for the X-ray
ones, 'ir'  for the infrared ones, or just the names of used lists
separated with comma, e.g. 'catalogs NVSS,FIRST,WISH' (the full
list of the CATS catalogs can be obtained from the CATS web page
http://cats.sao.ru/doc/CATS\_list.html). The keyword 'epoch' sets the
equinox of the input coordinates. 'Flux' sets flux limits.

The second example of the e-mail task is the matching procedure. There
are some additional keywords 'window box x=30$''$  y=10$'$, where 'box' is
the type of matching window (others are 'circle' and 'ellipse'), 'x' and
'y' are along longitude  and latitude of the corresponding coordinate
system, respectively. Note that 'x' and 'y' are half the side lengths
of the search 'box,' the value following the 'circle' keyword is its
radius, and those following the 'ellipse' keyword are the semi-major
and semi-minor axes of the ellipse.

\bigskip
{\large\bf 4~~~~~~~THE MAIN CATALOGS}

The major source of the radio catalogs in CATS is the collection one
of us (Andernach, 1990, 1999), who has spent considerable effort to
recover older source lists not previously available in electronic form,
using a scanner and optical character recognition software (Andernach,
2009). This collection is complemented in several ways: contributions
from authors, astro-ph preprints, tables from electronic journals and
the CDS catalog archive, as well as occasional manual retyping of the
original catalog. About 70 catalogs have been typed and/or corrected
manually at SAO RAS by S.Trushkin. The largest catalogs (e.g. NVSS,
FIRST, SDSS, etc.) were copied from web sites of the catalog authors.

CATS is mainly a radio-astronomical catalog service. All major catalogs
incorporated into CATS are shown in Tables 1 (adapted and updated from
Table~1 of Andernach, 1999) and 2.

{\bf Table~1.} Major radio astronomical catalogs available in CATS \\[-3ex]
\renewcommand{\baselinestretch}{.85}
{\small
\begin{verbatim}
 Freq. Name  Year of  RA (h)    Dec (deg) Half Power  S_min Number of
 (MHz)         publ.   l (d)      b (d)   BeamWidth/' (mJy)  objects
---------------------------------------------------------------------
10-25 UTR-2  78-02    0-24       > -13       25..60   10000    2237
  38  8C     90/95    0-24       > +60         4.5    1000     5859
  74  VLSS  2004-07   0-24       >-30          1.3     350    68311
  80  CUL1    73      0-24       -48,+35       3.7    2000      999
  80  CUL2    75      0-24       -48,+35       3.7    2000     1748
  82  IPS     87      0-24       -10,+83     27x350    500     1789
 151  6CI     85      0-24        < +80        4.5     200     1761
 151  6CII    88     8.5-17.5    +30,+51       4.5     200     8278
 151  6CIII   90     5.5-18.3    +48,+68       4.5     200     8749
 151  6CIV    91      0-24       +67,+82       4.5     200     5421
 151  6CVa    93     1.6- 6.2    +48,+68       4.5    ~300     2229
 151  6CVb    93    17.3-20.4    +48,+68       4.5    ~300     1229
 151  6CVI    93    22.6- 9.1    +30,+51       4.5    ~300     6752
 151  7CI     90    (10.5+41)    (6.5+45)      1.2      80     4723
 151  7CII    95      15-19      +54,+76       1.2    ~100     2702
 151  7CIII   96      9-16       +20,+35       1.2    ~150     5526
 160  CUL3    77      0-24       -48,+35       1.9    1200     2045
 178  4C      65      0-24        -7,+80     15x7.5   2000     4844
 232  MIYUN   96      0-24       +30,+90       3.8    ~100    34426
 325  WENSS   98      0-24       +30,+90       0.9      18   229420
 327  WSRT    91     5 fields   (+40,+72)     ~1.0       3     4157
 352  WISH   2002     0-24       -25,-9        0.9      18    90357
 365  TXS     96      0-24     -35.5,+71.5     ~.1     250    66841
 408  MRC    81/91    0-24      -85,+18.5      ~3      700    12141
 408  B2     70-73    0-24       +24,+40      3 x10    250     9929
 408  B3      85      0-24       +37,+47      3 x 5    100    13354
 608  WSRT    91    sev.fields  (~40,~72)      0.5       3     1693
 611  NAIC    75      22-13      -3,+19       12.6     350     3122
 843  SUMSS  99-06    0-24        <-30         0.72      6   210412
1400  GB      72      7-16       +46,+52      10x11     90     1086
1400  GB2     78      7-17       +32,+40      10x11     90     2022
1400  WB92    92      0-24       -5,+82       10x11   ~150    31524
1400  NVSS    98      0-24       -40,+90       0.9     2.0  1810668
1400  FIRST  98-07    7.3,17.4   22.2,57.6     0.1     1.0   816331
             98-07   21.3,3.3   -11.5,+1.6
1400  PDF     98    1.1 -1.3     -46,-45     0.1-0.2   0.1     1079
1500  VLANEP  94    17.4,18.5   63.6,70.4     0.25     0.5     2436
2700  PKS    (90)     0-24       -90,+27       ~8       50     8264
3900  Z       89      0-24        0,+14      1.2x52     50     8503
3900  RC     91/93    0-24       4.5,5.5     1.2x52      4     1189
3900  Z2      95      0-24        0,+14      1.2x52     40     2943
4850  MG1-4  86-91    var.        0,+39       ~3.5      50    24180
4850  87GB    91      0-24        0,+75       ~3.5      25    54579
4850  GB6     96      0-24        0,+75       ~3.5      18    75162
4850  PMNM    94      0-24       -88,-37       4.9      25    15045
4850  PMN-S   94      0-24      -87.5,-37      4.2      20    23277
4850  PMN-T   94      0-24       -29,-9.5      4.2      42    13363
4850  PMN-E   95      0-24       -9.5,+10      4.2      42    11774
4850  PMN-Z   96      0-24       -37,-29       4.2      70     2400
---------------- Catalogs covering the Galactic Plane:  ------------
  31  NEK      88   350<l<250    |b|<2.5      13x 11  4000      703
 151  7C(G)    98    80<l<180    |b|<5.5       1.2    ~100     6262
 327  WSRTGP   96    43<l<91     |b|<1.6      ~1.0     ~10     3984
 843  MGPS-2  07    245<l<365    |b|<10        0.7      10    48850
1400  GPSR     90    20<l<120    |b|<0.8       0.08     25     1992
1408  RRF      90   357<l<95.5   |b|<4.0       9.4      98      884
1420  RRF      98   95.5<l<240   -4<b<+5       9.4      80     1830
1400  GPSR     92   350<l<40     |b|<1.8      0.08      25     1457
2700  F3R      90   357<l<240    |b|< 5        4.3      40     6483
4875  ADP79    79   357<l<60     |b|<1         2.6    ~120     1186
5000  GT       86    40<l<220    |b|<2         2.8      70     1274
5000  GPSR     94   350<l<40     |b|<0.4      ~0.07      3     1272
5000  GPSR     79   190<l<40     |b|<2         4.1     260      915
\end{verbatim}
}

{\bf Table~2.} Some catalogs of other wavelength ranges in CATS \\[-3ex]
{\small
\begin{verbatim}
lambda    Name    PublYear    RA        Dec         N(objects)
--------------------------------------------------------------
opt       PGC        89      0-24     -90,+90        73197
opt       MCG        75      0-24    -33.5,+90       31886
opt       MSL        85      0-24     -90,+90       181603
opt       SDSS DR6  2007    several                 216906 Gals
                             fields                  22033 QSOs
ir       IRASPSC     87      0-24     -90,+90       245889
ir       IRASFSC     89      0-24     |b|>10        235935
ir       IRASSSC     89      0-24     -90,+90        43886
ir        2MASS     2000     0-24     -90,+90    470992970
Xray      ROSAT      95      0-24     -90,+90        74301
mix      QSO HB2     93      0-24     -84,+86         7315
mix      VERON+11    93      0-24     -83,+85        48921
\end{verbatim}
}

\renewcommand{\baselinestretch}{1.0}

CATS also contains observational and combined catalogs of various Galactic
and extragalactic objects unified in the combined tables described in
Verkhodanov et al.\ (2005) where the references to all the mentioned
catalogs can be found.

\bigskip
{\large\bf 5~~~~~~~~SUMMARY}

CATS provides a simple and convenient access to astrophysical data and
complements the data available from other services, most notably for
radio continuum flux densities, for which it provides the largest amount
of data in any such service.  Operation with CATS permits astronomers
to search for peculiar objects and study physical processes in sources
of cosmic radiation. 

Until October 2004, we registered over 28000 requests for the 'select' or
'match' procedures, which are the most popular. The most popular catalogs
for FTP-copying over the last five years are QSO by Hewitt and Burbidge
(20 times), PGC (19 times), and NVSS (18 times). CATS processes daily up
to 1000 HTTP-requests for information concerning the catalog descriptions.

CATS is being expanded continuously and presently comprises more than 400
catalogs including all the RATAN-600 catalogs.  In its present form CATS
occupies about 60Gb of disk space, which makes its installation possible
also on notebooks. One of the possible future developments of the system
is the preparation of the VOTable format for output tables. At present,
limited resources preclude linkage with the Virtual Observatory (VO).

\bigskip
{\large\bf 6~~~~~~~~REFERENCES}

Andernach, H. (1990) The Need for a Radio Astronomical Data Base: First
Results of a Campaign. {\it Bull.\ Inf.\ CDS} ~38, 69.

Andernach, H. (1994) How Complete is the Electronic Archive at CDS
for information on Extragalactic Objects\,? ~{\it Bull.\ Inf.\ CDS} ~45, 35,
{\tts astro-ph/9411027}

Andernach, H. (1999) in Internet Resources for Professional Astronomy,
Lecture Notes of a course given at {\it IX~Canary Islands Winter School of
Astrophysics}, in {\it Astrophysics with Large Database in the Internet Age},
Nov. 1997, eds. M. Kidger, I. Perez-Fournon, F. Sanchez, Cambridge
Univ. Press, p.\ 67 ({\tts astro-ph/9807346})

Andernach, H. (2009) Safeguarding Old and New Journal Tables for the VO:
Status for Extragalactic and Radio Data, {\it Data Science Journal}, vol.\ 8,
in press ~({\tts http://dsj.codataweb.org}), {\tts arXiv:0901.2805}

Egret, D. (1983) SIMBAD Story - a Description of the Database of the
Strasbourg Stellar Data Center. {\it Bull.\ Inf.\ CDS} ~24, 109.

Helou, G., Madore, B.F., Bicay, M.D., Schmitz, M., \& Liang, J. (1990)
in Windows on Galaxies, Proc.\ 6th Workshop of the {\it Advanced
School of Astronomy of the Ettore Majorana Centre}, Erice, May 21-23,
1989, Dordrecht: Kluwer, 1990, eds. G. Fabbiano, J.S. Gallagher, \&
A.\ Renzini, {\it Astrophys.\ \& Space Sci.\ Library}, 160, 109

Mazzarella, J.M., Madore, B.F., Bennett, J., Corwin, H., Helou, G.,
Kelly, A.,Schmitz, M., \& Skiff. B. (2002) Analysis and Visualization
of Multiwavelength Spectral Energy Distributions in the NASA/IPAC
Extragalactic Database (NED). {\it Astronomical Data Analysis~II},
eds.\ J.-L.Starck, F.D. Murtagh, Proc.\ of the SPIE, 4847, 254.

Ochsenbein, F., Bauer P., \& Marcout, J. (2000) The VizieR database of
astronomical catalogues. {\it Astron.\ \& Astrophys.\ Suppl.\ Ser.} 143, 23--32.

Paturel, G., Andernach, H., Bottinelli, L., Di Nella, H., Durand, N.,
Garnier, R., Gouguenheim, L., Lanoix, P., Marthinet, M.-C., Petit,
C., Rousseau,  J., Theureau, G., \& Vauglin, I. (1997) Extragalactic
database. VII. Reduction of astrophysical parameters. {\it Astron.\ \&
Astrophys.\ Suppl.\ Ser.} 124, 109--122.

Trushkin, S.A., Verkhodanov, O.V., Chernenkov, V.N., \& Andernach,
H. (2000) CATS - The Largest Radio Astronomical Database: Galactic
Facilities. {\it Baltic Astronomy} 9, 608.

Verkhodanov, O.V. \& Trushkin, S.A. (1994) Report 228 SAO RAS, Library
SAO RAS, Nizhnij Arkhys, Russia.

Verkhodanov, O.V. \& Trushkin, S.A. (1995a) Preprint No 106 SAO RAS,
Nizhnij Arkhyz, Library SAO, Russia, 66

Verkhodanov, O.V. \& Trushkin, S.A. (1995b) in: Proc.\ {\it XXVI Radio
Astronomy Conference}, St.\,Petersburg, {\it IAA RAS}, 252.

Verkhodanov, O.V., Trushkin, S.A., Andernach, H., \& Chernenkov,
V.N. (1997) The CATS database to operate with astrophysical catalogs,
in {\it Astronomical Data Analysis Software and Systems~VI}, eds.:
G.\ Hunt \& H.E.\ Payne (Eds.), {\it ASP Conf.\ Ser.} ~125, 322, astro-ph/9610262.

Verkhodanov, O.V., Trushkin, S.A., Chernenkov, V.N., \& Andernach,
H. (2000) CATS - The Largest Radio Astronomical Database: Extragalactic
Facilities. {\it Baltic Astronomy} ~9, 604.

Verkhodanov, O.V., Trushkin, S.A., Andernach, H., \& Chernenkov,
V.N. (2005) Current status of the CATS database. {\it Bull.\ SAO} ~58,
118--129, {\tts astro-ph/0705.2959}

Wenger, M., Ochsenbein, F., Egret, D., Dubois, P., Bonnarel, F.,
Borde, S., Genova, F., Jasniewicz,G., Laloe,S., Lesteven, S., \& Monier,
R. (2000) The SIMBAD astronomical database. The CDS reference database
for astronomical objects. {\it Astron.\ \& Astrophys.\ Suppl.\ Ser.} 143, 9,
{\tts astro-ph/0002110}.

\end{document}